\documentclass[
twocolumn,
pra,showpacs,preprintnumbers,amsmath,amssymb]{revtex4}

\usepackage{graphicx}
\usepackage{amssymb}
\usepackage{amsmath}
\usepackage{dsfont}
\usepackage{bm}
\usepackage{mathrsfs}
\usepackage{times}

\begin{document}
\title{An alternative calculation of the Casimir forces between birefringent plates}
\author{T.\ G.\ Philbin$^1$ and U.\ Leonhardt$^2$}
\affiliation{$^1$Max Planck Research Group, Institute of Optics, Information and Photonics, G\"{u}nther-Scharowsky-Str.\ 1/ Bau 24, 91058 Erlangen, Germany \\
$^2$School of Physics and Astronomy, University of St Andrews,
North Haugh, St Andrews, Fife, KY16 9SS, Scotland}

\begin{abstract}
Barash has calculated the Casimir forces between parallel birefringent plates with optical axes parallel to the plate boundaries [Izv.\ Vyssh.\ Uchebn.\ Zaved., Radiofiz., {\bf 12}, 1637 (1978)]. The interesting new feature of the solution compared to the case of isotropic plates is the existence of a Casimir torque which acts to line up the optical axes if they are not parallel or perpendicular. The forces were found from a calculation of the Helmholtz free energy of the electromagnetic field. Given the length of the calculations in this problem and hopes of an experimental measurement of the torque, it is important to check the results for the Casimir forces by a different method. We provide this check by calculating the electromagnetic stress tensor between the plates and showing that the resulting forces are in agreement with those found by Barash.
\hspace*{\fill}
\end{abstract}
\date{\today}

\maketitle

\section{Introduction}
Not long after Casimir's landmark paper~\cite{cas48} on the attractive force between two parallel perfect mirrors, Lifshitz generalized the result to the case of isotropic dielectric plates~\cite{lif55,LL}. Lifshitz's formidable calculation was subsequently extended to allow for a third dielectric between the plates~\cite{dzy61}, the case in mind being that of parallel plates immersed in a fluid. An obvious further generalization of the problem is to allow for anisotropic permittivities in the plates. The first exact solution for anisotropic plates was found in 1978 by Barash~\cite{bar78}, who considered uniaxial (birefringent) plates with optical axes parallel to the plate boundaries, separated by an isotropic dielectric. An expression was found for the Helmholtz free energy of the electromagnetic field at finite temperature, which was a function of the angle between the optical axes of the plates as well as of the plate separation. This means that there is a torque on the plates as well as a perpendicular force. These forces persist even at zero temperature due to the zero point energy of the quantum vacuum. The anisotropy thus leads to a new phenomenon in Casimir forces, a torque that acts to align the optical axes of the plates if they are not parallel or perpendicular~\cite{bar78,mun05}. These results have been used to propose an experimental measurement of the Casimir torque using birefringent crystals immersed in ethanol~\cite{mun05}.

The degree of complexity of this problem is considerable. This is reflected in the very lengthy expression found in~\cite{bar78} for the free energy, which we will not reproduce here (it can also be found in~\cite{mun05}, but note an important misprint~\footnote{\label{barcom}In the function $\gamma$ which appears in the free energy in ~\cite{bar78} and~\cite{mun05}, the denominator of the last term should have a minus sign, instead of a product, between the two squared quantities. We are indebted to Y.\ S.\ Barash for confirming this misprint.}). Given the length of the calculations and plans for an experiment to test the theory, it is highly desirable to have an independent solution of the problem using a different method. To date however, the only other analysis of this problem was a very simplified calculation~\cite{van95} which gave a rough estimate of the Casimir forces. (The simpler problem of one isotropic and one birefringent plate with optical axis perpendicular to the boundary was analyzed recently in~\cite{ros08}.) 

In this paper we solve the problem exactly using Lifshitz's approach~\cite{lif55,dzy61,LL}, in which the electromagnetic stress tensor is found by a Green-function method. At room temperature, the forces for a realistic experimental set-up were found in~\cite{mun05} to be determined by the zero-temperature Casimir contribution, with contributions from thermal radiation being negligible. We therefore ignore thermal effects and find the exact solution at zero temperature. Our approach has considerable methodological interest, since we present a much simpler, more physical route to the stress tensor than is found in the standard Lifshitz theory~\cite{lif55,dzy61,LL}. We developed this formalism in an analysis of the Casimir forces on moving plates~\cite{phi}; here its general usefulness is demonstrated, since the solution obtained in this paper would have required considerably more labour if we had taken the usual approach~\cite{lif55,dzy61,LL}. Our treatment also generalizes the problem to allow for a magnetic response in the isotropic medium between the plates. Until recently an influence of magnetic permeability on Casimir forces was considered of only theoretical interest, but the development of metamaterials with engineered magnetic responses raises new possibilities~\cite{hen05,lev,pir08,ros08}.  

From the electromagnetic stress tensor between the plates we obtain an expression for the electromagnetic energy; when there is no magnetic response our result agrees with the low-temperature limit of Barash's formula. This agreement is highly non-trivial: the formula for the energy in both solutions requires integrations and the integrands in each case are completely different; numerical evaluation of the integrals shows that the two expressions give the same number.

Section~\ref{sec:notation} sets out the problem and notation in detail. In Sec.~\ref{sec:theory} we show how the electromagnetic stress tensor can be found from the Green tensor of the vector potential. The Green tensor is calculated in Sec.~\ref{sec:green}, by a novel method, and in Sec.~\ref{sec:stress} we present the stress tensor and electromagnetic energy.

\section{\label{sec:notation}Notation and geometry}
\begin{figure}
\includegraphics[width=15.0pc]{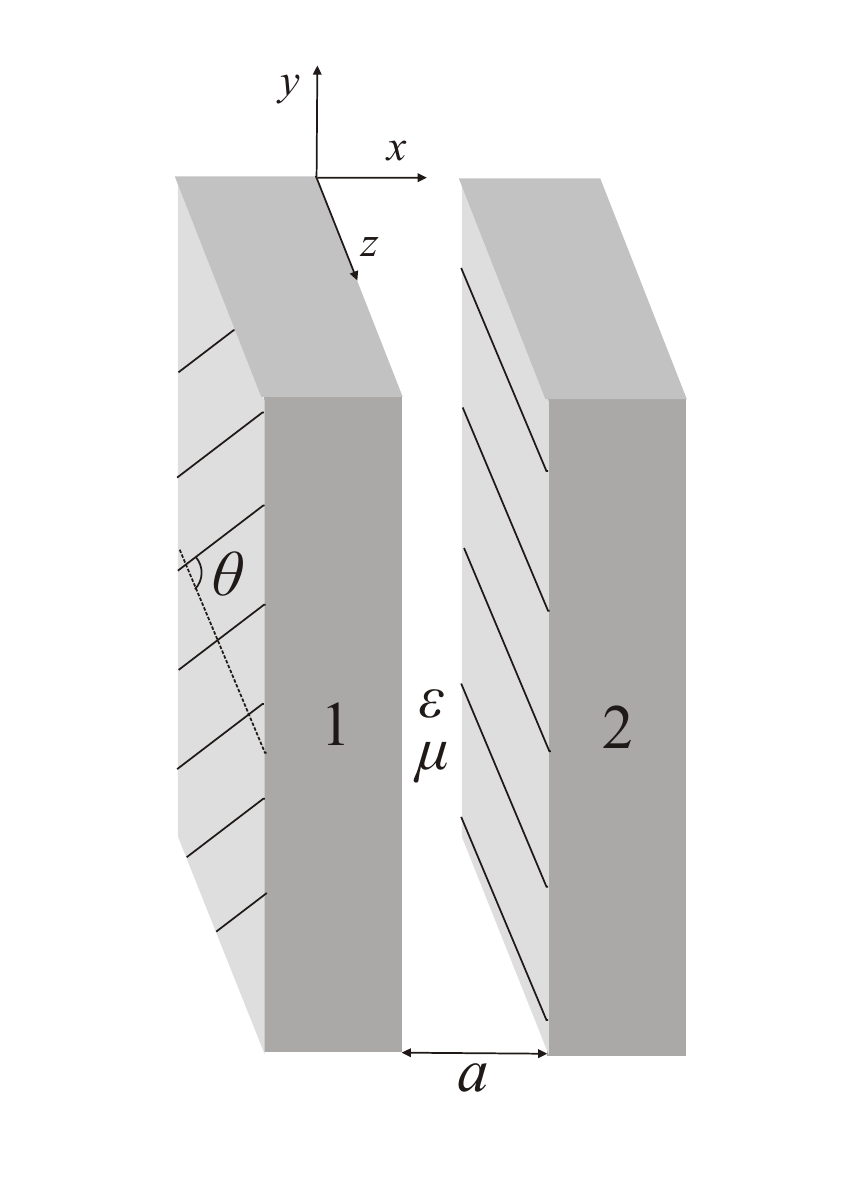}
\vspace*{-7mm}
\caption{Two plates birefringent plates lie in the $yz$-plane with constant separation $a$. The lines on the plates represent the directions of the optical axes, which also lie in the $yz$-plane. The axis of Plate~2 lies in the $z$-direction while that of Plate~1 is rotated relative to this by $\theta$. Between the plates is an isotropic medium of permittivity $\epsilon$ and permeability $\mu$.
 \label{fig1}}
\end{figure}

We consider the arrangement depicted in Fig.~\ref{fig1}. Two birefringent plates lie in the $yz$-plane, separated by a distance $a$. The optical axis of Plate~1 lies at an angle $\theta$ to the $z$-axis while the optical axis of Plate~2 is directed along this axis. Between the plates there is an isotropic material with frequency-dependent permittivity $\epsilon$ and permeability $\mu$. 

The permittivity along the optical axis is denoted by $ \epsilon_{{\scriptscriptstyle\parallel}}$ and the permitivity in directions perpendicular to the optical axis by  $\epsilon_{{\scriptscriptstyle\perp}}$; these are also frequency dependent but to reduce the notational clutter we will not make this explicit. It follows that the dielectric tensors~\cite{born,LLcm} of Plates~1 and~2 are, respectively,
\begin{gather}
\bm{\Lambda}\left( \begin{array}{ccc}
\epsilon_{1{\scriptscriptstyle\perp}} & 0 & 0 \\
0 & \epsilon_{1{\scriptscriptstyle\perp}} & 0 \\
0 & 0 & \epsilon_{1{\scriptscriptstyle\parallel}} 
\end{array}\right)\bm{\Lambda}^T,  \label{perm1} \displaybreak[0] \\[5pt]
\left( \begin{array}{ccc}
\epsilon_{2{\scriptscriptstyle\perp}} & 0 & 0 \\
0 & \epsilon_{2{\scriptscriptstyle\perp}} & 0 \\
0 & 0 & \epsilon_{2{\scriptscriptstyle\parallel}} 
\end{array}\right),  \label{perm2}
\end{gather}
where
\begin{equation} \label{lambda}
 \bm{\Lambda}=\left( \begin{array}{ccc}
1 & 0 & 0 \\
0 & \cos\theta & \sin\theta \\
0 & -\sin\theta & \cos\theta 
\end{array}\right)
\end{equation}
is a rotation matrix about the $x$-axis. The angle $\theta$ in (\ref{lambda}) produces the orientation of Plate~1 portrayed in Fig.~\ref{fig1}.

Luckily, the principle axes of a uniaxial crystal are frequency independent~\cite{born} so the permittivity  tensors are given by (\ref{perm1})--(\ref{lambda}) at all frequencies. This property is crucial in making the problem tractable, although this was not explicitly pointed out in~\cite{bar78}. It does not hold for biaxial crystals (different permittivities in all three principle directions)---``dispersion of the axes"~\cite{born}.

\section{\label{sec:theory}Lifshitz theory}
The Casimir forces on the plates are determined by the expectation value of the electromagnetic stress tensor in the medium between the plates~\cite{dzy61}. This expectation value is computed for the zero-temperature ground state of the electromagnetic field, and the remarkable fact is that it is non-zero even though there are no electromagnetic fields present. The origin of this stress is the quantum zero-point energy of the electromagnetic field, which is modified by the materials~\cite{milonni}. From the classical Lagrangian of macroscopic electromagnetism
\begin{equation} \label{emlag}
L=\int d^3x\,\frac{1}{2}(\mathbf{D}\cdot\mathbf{E}-\mathbf{H}\cdot\mathbf{B})
\end{equation}
one easily obtains the expression for the stress tensor by the usual methods of field theory~\cite{bar}. The required quantity is the expectation value of the quantum version of this stress tensor:
\begin{equation} \label{stress}
\begin{split}
\bm{\sigma}=\,&\varepsilon_0\langle \mathbf{\hat{D}}\otimes\mathbf{\hat{E}}\rangle+\mu_0^{-1}\langle \mathbf{\hat{H}}\otimes\mathbf{\hat{B}}\rangle \\
&-\frac{1}{2}\mathds{1}(\varepsilon_0\langle \mathbf{\hat{D}}\cdot\mathbf{\hat{E}}\rangle+\mu_0^{-1}\langle \mathbf{\hat{H}}\cdot\mathbf{\hat{B}}\rangle).
\end{split}
\end{equation}
The methodology of Lifshitz theory is to compute the expectation values in the stress (\ref{stress}) using Green tensors of the vector potential $\mathbf{A}$ (with the scalar potential set to zero)~\cite{LL}. In quantum electrodynamics the basic computational quantity is the Feynman propagator or Green tensor $\mathbf{G}^F$, given by~\cite{LL}
\begin{equation} \label{fgreen}
\langle T\mathbf{\hat{A}}(\mathbf{r},t)\otimes\mathbf{\hat{A}}(\mathbf{r'},t')\rangle=-\frac{i\hbar\mu_0}{2\pi}\mathbf{G}^F(\mathbf{r},t;\mathbf{r'},t').
\end{equation}
This expression can be used to calculate field correlation functions which in turn determine the stress (\ref{stress}) (see below). At finite temperature it can be shown~\cite{dzy61,LL}, through use of the fluctuation-dissipation theorem among other things, that the required correlation functions have a simple relation to the retarded Green tensor. In the zero-temperature case, however, one can obtain this result much more simply using the fundamental starting point (\ref{fgreen}), without having to invoke the fluctuation-dissipation theorem. We briefly describe the derivation.

A Green tensor for the vector-potential wave equation satisfies
\begin{gather}
\mathbf{G}(\mathbf{r},t,\mathbf{r'},t')=\int^\infty_{-\infty}d\omega \, \mathbf{G}(\mathbf{r},\mathbf{r'},\omega)\,e^{-i\omega(t-t')},  \label{freqG} \\[5pt]
\left(\nabla\times\frac{1}{\mu(\omega)}\nabla\times-\frac{\omega^2}{c^2}\epsilon(\omega)\right)\mathbf{G}(\mathbf{r},\mathbf{r'},\omega)=\mathds{1}\delta(\mathbf{r}-\mathbf{r'}), \label{green}
\end{gather}
and depends on the material boundary conditions. From the monochromatic Maxwell equations leading to (\ref{green}) one deduces the following property of the monochromatic Green tensor in (\ref{freqG})--(\ref{green}):
\begin{equation}  \label{real}
\mathbf{G}(\mathbf{r},\mathbf{r'},-\omega^*)=\mathbf{G}^*(\mathbf{r},\mathbf{r'},\omega).
\end{equation}
A complete set of boundary conditions for (\ref{green}) requires not only the spatial boundary conditions given by the plates, but also conditions in time. One thus has the usual choice~\cite{jac} of retarded or advanced boundary conditions, or something more complicated. The Feynman Green tensor (\ref{fgreen}) can be constructed from the retarded ($\mathbf{G}^R$) and advanced ($\mathbf{G}^A$) Green tensors, which are easier to calculate. In the limit $\mathbf{r}\rightarrow\mathbf{r'}$, $t\rightarrow t'$, which is required to construct the stress tensor (\ref{stress}) from the matrix element (\ref{fgreen}), the relationship is~\cite{bar}
\begin{equation} \label{greenrel}
\begin{split}
\lim_{\stackrel{\mathbf{r}\rightarrow\mathbf{r'}}{t\rightarrow t'}}&\mathbf{G}^F(\mathbf{r},t,\mathbf{r'},t') \\
  &=\frac{1}{2}\lim_{\stackrel{\mathbf{r}\rightarrow\mathbf{r'}}{t\rightarrow t'}}\left(\mathbf{G}^R(\mathbf{r},t,\mathbf{r'},t')+\mathbf{G}^A(\mathbf{r},t,\mathbf{r'},t')\right).
\end{split}
\end{equation}
Since the boundary conditions are time symmetric, the advanced Green tensor is found by time reversing the retarded case: 
\begin{equation} \label{retadv}
\mathbf{G}^A(\mathbf{r},t,\mathbf{r'},t')=\mathbf{G}^R(\mathbf{r},-t,\mathbf{r'},-t')
\end{equation}
It follows from (\ref{greenrel}) and (\ref{retadv}) that
\begin{equation} \label{feynret}
\lim_{\stackrel{\mathbf{r}\rightarrow\mathbf{r'}}{t\rightarrow t'}}\mathbf{G}^F(\mathbf{r},t,\mathbf{r'},t')
  =\lim_{\stackrel{\mathbf{r}\rightarrow\mathbf{r'}}{t\rightarrow t'}}\mathbf{G}^R(\mathbf{r},t,\mathbf{r'},t'),
\end{equation}
so that only the retarded Green tensor is required to compute the stress.
Once the retarded Green tensor is at hand the quadratic expectation values of the electric and magnetic fields in the stress tensor (\ref{stress}) are found from~\cite{LL}
\begin{gather}
\langle\mathbf{\hat{D}}(\mathbf{r})\!\otimes\!\mathbf{\hat{E}}(\mathbf{r'})\rangle=-\frac{\hbar\mu_0}{\pi}\int_0^\infty d\xi\,\epsilon(i\xi)\xi^2\mathbf{G}^R(\mathbf{r},\mathbf{r'},i\xi), \label{EE} \\
\begin{split}
\langle\mathbf{\hat{H}}(\mathbf{r})&\,\!\otimes\!\mathbf{\hat{B}}(\mathbf{r'})\rangle \\[5pt]
&=\frac{\hbar\mu_0}{\pi}\int_0^\infty d\xi\,\frac{1}{\mu(i\xi)}\nabla\times\mathbf{G}^R(\mathbf{r},\mathbf{r'},i\xi)\times\stackrel{\leftarrow}{\nabla'}. \label{BB}
\end{split}
\end{gather}
These follow from (\ref{fgreen}), (\ref{feynret}) and (\ref{freqG}), with a final Wick rotation to imaginary frequencies ($\omega=i\xi$) in which (\ref{real}) is used. The switch to imaginary frequncies ensures that the integrals are well behaved~\cite{lif55,LL,dzy61}.

\section{\label{sec:green}Green tensor}
The problem has now been reduced to that of finding the classical retarded Green tensor for the vector potential in the medium between the plates.
With retarded boundary conditions, the solution of the mono\-chromatic equation (\ref{green}) has a simple physical meaning: an oscillating dipole at the point $\mathbf{r'}$ emits electromagnetic waves of frequency $\omega$ and $\mathbf{G}^R(\mathbf{r},\mathbf{r'},\omega)$ is the resulting vector potential at the point $\mathbf{r}$. The second index in $G^R_{ij}$ represents the orientation of the dipole at $\mathbf{r'}$, while the first index represents the components of the vector potential at $\mathbf{r}$. Using this physical consideration it is clear from Fig.~\ref{fig1} that the solution will be a linear superposition of waves that have reflected off the plates, with the number of reflections ranging from zero to infinity. To write down the solution we Fourier transform the Green tensor
\begin{equation} \label{four}
\begin{split}
\mathbf{\widetilde{G}}^R&(x,x',u,v,i\xi) \\
&=\int_{-\infty}^\infty dy\int_{-\infty}^\infty dz\,\mathbf{G}^R(\mathbf{r},\mathbf{r'},i\xi)\,e^{-iu(y-y')-iv(z-z')}
\end{split}
\end{equation}
so that we decompose the waves emitted by the dipole into plane waves. In the absence of the plates the solution is the bare Green tensor~\cite{dzy61,LL}
\begin{gather}
\mathbf{\widetilde{G}}^R_b(x,x',u,v,i\xi)=
\left\{\begin{array}{l} e^{-w(x-x')}\boldsymbol{\mathcal{G}_+}, \quad x>x'  \\[5pt] e^{w(x-x')}\boldsymbol{\mathcal{G}_-}, \quad x<x' \end{array}\right.,  \label{bare} \\[6pt]
\boldsymbol{\mathcal{G}_\pm}=-\frac{1}{2\epsilon w\kappa^2}\left[\left(\begin{array}{c} \pm w \\ iu \\ iv \end{array}\right)\!\!\otimes\!\!\left(\begin{array}{c} \pm w \\ iu \\ iv \end{array}\right)-\epsilon\mu\kappa^2\mathds{1}\right], \label{Gcal} \\[6pt]
 \kappa=\frac{\xi}{c}, \qquad w=\sqrt{u^2+v^2+\epsilon\mu\kappa^2}. \label{w}
\end{gather}
The two possibilities in (\ref{bare}) are linearly-polarized plane waves propagating to the right (first line) or to the left (second line), with wave vectors 
\begin{equation} 
\mathbf{k}_\pm =(\pm iw,u,v). \label{ks}
\end{equation}
The imaginary $x$-component of the wave vectors is a consequence of the imaginary frequency, and 
in (\ref{w}) we simply have the relation $\sqrt{\epsilon\mu}\,\omega=ck$. In physical terms the vacuum solution (\ref{bare}) is trivial: it is the only way the dipole can propagate plane waves from $x'$ to $x$. In the presence of the plates both plane waves in (\ref{bare}) will reflect off the plates and reverse direction, so the left-moving plane wave can propagate from $x'$ to $x$ even if $x>x'$, with similar considerations applying to the right-moving plane wave. Consequently, both the right- and left-moving waves will appear in the solution regardless of whether $x$ is greater or less than $x'$, in contrast to the vacuum solution (\ref{bare}). Let $\mathbf{R}_2$ be the reflection operator (matrix) that transforms a right-moving plane wave at plate~2 into the resulting reflected left-moving plane wave, and let $\mathbf{R}_1$ be the reflection operator that transforms a left-moving plane wave at plate~1 into the reflected right-moving plane wave. We can now write down the solution (functional dependences are suppressed):
\begin{align} 
\mathbf{\widetilde{G}}^R=\,&\mathbf{\widetilde{G}}^R_b-e^{-w(x-x')}\boldsymbol{\mathcal{G}_+}-e^{w(x-x')}\boldsymbol{\mathcal{G}_-} \nonumber \\
& +\left(\mathds{1}-e^{-2a}\mathbf{R}_1\mathbf{R}_2\right)^{-1} \nonumber \\
& \times  \left(e^{-w(x-x')}\boldsymbol{\mathcal{G}_+}+e^{-w(x+x')}\mathbf{R}_1\boldsymbol{\mathcal{G}_-}\right)\nonumber  \\
& +\left(\mathds{1}-e^{-2a}\mathbf{R}_2\mathbf{R}_1\right)^{-1} \nonumber \\
&  \times  \left(e^{w(x-x')}\boldsymbol{\mathcal{G}_-}+e^{w(x+x'-2a)}\mathbf{R}_2\boldsymbol{\mathcal{G}_+}\right). \label{soln}
\end{align}
The first line in (\ref{soln}) subtracts the left- or right-moving plane wave in (\ref{bare}), depending on whether $x$ is greater or less than $x'$. This subtraction is necessary because the direct propagation, without reflections, of both the right- and left-moving plane waves from $x'$ to $x$ is contained in the remaining terms in (\ref{soln}); but only one of these propagations is possible, depending on whether $x$ is greater or less than $x'$, and the first line in (\ref{soln}) automatically subtracts the irrelevant one. The inverse matrices in (\ref{soln}) are geometrical series representing every possible number of double reflections off both plates, the exponentials providing the propagation distance $2a$ for each double reflection. The initial right- and left-moving plane waves that leave $x'$ reach $x$ after both an even and odd number of reflections; this explains the terms multiplying the inverse matrices in (\ref{soln}). Each term in (\ref{soln}), after the series expansion of the inverse matrices, has  an overall exponential factor that accounts for the propagation distance involved, with $e^{-ws}$, $s>0$, representing a propagation distance $s$ to the right for the initial right-moving plane wave, but to the left for the initial left-moving plane wave.

The solution for the Green tensor, and thereby the Casimir forces, is thus determined once we compute the reflection operators $\mathbf{R}_1$ and $\mathbf{R}_2$, which describe the reflection of a linearly-polarized plane wave at each plate. This is a tedious business, but note that we have avoided the vastly more complicated task of having to solve the differential equation for the Green tensor with boundary conditions given by the plates. The use of physical reasoning to write down the solution (\ref{soln}) for the Green tensor is a novel feature of our approach to the Lifshitz theory for parallel plates.

\subsection{Reflection operator for Plate~2}
Consider first the reflection operator for Plate~2. There are two kinds of plane wave that can propagate in a birefringent crystal: the {\it ordinary wave} and the {\it extraordinary wave}~\cite{born,LLcm}. When the plane wave with wave vector $\mathbf{k}_+$ (see (\ref{ks})) impinges on Plate~2 it produces an ordinary and an extraordinary wave in the plate with wave vectors~\cite{born,LLcm}
\begin{gather}
\mathbf{k}_{2o}=(iw_{2o},u,v), \qquad w_{2o}=\sqrt{u^2+v^2+\epsilon_{2{\scriptscriptstyle\perp}}\kappa^2}, \label{k2o} \\
\mathbf{k}_{2e}=(iw_{2e},u,v), \quad w_{2e}=\sqrt{u^2+v^2\epsilon_{2{\scriptscriptstyle\parallel}}/\epsilon_{2{\scriptscriptstyle\perp}}+\epsilon_{2{\scriptscriptstyle\parallel}}\kappa^2}. \label{k2e}
\end{gather}

Although we are dealing with vector-potential waves, the electric field is proportional to $\mathbf{A}$, so that $\mathbf{A}$ oscillates in the polarization direction.
The vector potential of the ordinary and extraordinary waves can be written
\begin{gather}
\mathbf{A}_{2o}=A_{2o} \left(\begin{array}{c} u \\ -iw_{2o} \\ 0 \end{array}\right)e^{i\mathbf{k}_{2o}\cdot\mathbf{r}+c\kappa t},  \label{opol} \\ 
\mathbf{A}_{2e}=A_{2e} \left(\begin{array}{c} iw_{2e}v \\ uv \\ v^2+\epsilon_{2{\scriptscriptstyle\perp}}\kappa^2 \end{array}\right)e^{i\mathbf{k}_{2e}\cdot\mathbf{r}+c\kappa t}, \label{epol}
\end{gather}
which reveals their polarizations. Note that the ordinary wave is always polarized perpendicular to the optical axis.

The reflected wave at Plate¬2 has wave vector $\mathbf{k}_-$ (see (\ref{ks})), but its polarization is rotated relative to the incident wave. This rotation of the polarization by the plates is the reason why the final answer for the Casimir energy will be so much more complicated than in the isotropic case. 

To find the reflection operator for the $\mathbf{k}_+$ plane wave we decompose it into its component with polarization perpendicular to the plane of incidence ($E$-polarization) and its component in this plane ($B$-polarization). The $E$- and $B$-polarization directions are given by the unit vectors
\begin{gather}
\mathbf{n}_{E}=\frac{1}{\sqrt{u^2+v^2}}\left(\begin{array}{c} 0 \\ -v \\ u \end{array}\right), \label{nE} \\[7pt]
\mathbf{n}_{B2}=\frac{1}{\kappa\sqrt{\epsilon\mu}\,\sqrt{u^2+v^2}}\left(\begin{array}{c} i(u^2+v^2) \\ uw \\ vw \end{array}\right). \label{nB2}
\end{gather}
The reason why there is no number subscript on $\mathbf{n}_{E}$ is that this is also the $E$-polarization direction for the $\mathbf{k}_-$ plane wave reflecting off Plate~1. We must now solve the reflection problem separately for the $E$- and $B$-polarizations. This is done as in the isotropic case by imposing the standard boundary conditions that arise from the macroscopic Maxwell equations~\cite{jac}. From (\ref{nE}) and (\ref{nB2}) we can take the incident $E$- and $B$-polarized waves to be
\begin{gather}
\mathbf{A}_{E}=A_{E} \left(\begin{array}{c}  0 \\ -v \\ u \end{array}\right)e^{i\mathbf{k}_{+}\cdot\mathbf{r}+c\kappa t}  \label{AE} \\ 
\mathbf{A}_{B}=A_{B} \left(\begin{array}{c} i(u^2+v^2) \\ uw \\ vw \end{array}\right)e^{i\mathbf{k}_{+}\cdot\mathbf{r}+c\kappa t}. \label{AB}
\end{gather}
The answer for the reflection cannot be expressed in terms of a scalar reflection coefficient because of the rotation of the polarization referred to above. Nevertheless, for both incident polarizations (\ref{AE}) and (\ref{AB}), everything can be written in terms of two scalars which refer to the transmitted ordinary and extraordinary waves. The first of these scalars is the ratio $A_{2e}/A_{2o}$; for the $E$-polarization we denote this ratio by $\alpha_2$, and for the $B$-polarization we denote it by $\gamma_2$:
\begin{gather} 
\alpha_2=\frac{iu(\epsilon w_{2o}+\epsilon_{2{\scriptscriptstyle\perp}}w)}{v(\epsilon w_{2o}^2+\epsilon_{2{\scriptscriptstyle\perp}}w_{2e}w)},  \label{al2} \\
\gamma_2=-\frac{ivw_{2o}(w+\mu w_{2o})}{\epsilon_{2{\scriptscriptstyle\perp}}u\kappa^2(w+\mu w_{2e})}.  \label{ga2}
\end{gather}
The second scalar relates the transmitted ordinary wave to the incident wave; for the $E$-polarization it is $A_{2o}/A_{E}$, which we denote by $\beta_2$, and for the $B$-polarization it is $A_{2o}/A_{B}$, which we denote it by $\delta_2$:
\begin{widetext}
\begin{equation}
\beta_2=\frac{2i\epsilon vw}{\epsilon w_{2o}w+\epsilon_{2{\scriptscriptstyle\perp}}u^2+iuv\alpha_2(\epsilon w+\epsilon_{2{\scriptscriptstyle\perp}}w_{2e})+\epsilon\mu(v^2+\epsilon_{2{\scriptscriptstyle\perp}}\kappa^2)}, \qquad
\delta_2=\frac{2\epsilon w(u^2+v^2)}{\epsilon uw_{2o}+i\epsilon\gamma_2vw_{2o}^2+\epsilon_{2{\scriptscriptstyle\perp}}w(u+i\gamma_2vw_{2e})}. \label{be2}
\end{equation}
\end{widetext}
The reflected waves for the two incident polarizations (\ref{AE}) and (\ref{AB}) are, respectively,
\begin{gather}
-\sqrt{u^2+v^2}(\mathbf{n}_E+\mathbf{m}_{E2})A_E\,e^{i\mathbf{k}_{-}\cdot\mathbf{r}+c\kappa t}, \\
-\kappa\sqrt{\epsilon\mu}\,\sqrt{u^2+v^2}(\mathbf{n}_{B2}+\mathbf{m}_{B2})A_B\,e^{i\mathbf{k}_{-}\cdot\mathbf{r}+c\kappa t},
\end{gather}
where the vectors $\mathbf{m}_{E2}$ and $\mathbf{m}_{B2}$ are
\begin{gather}
\mathbf{m}_{E2}=\frac{1}{\sqrt{u^2+v^2}}\left(\begin{array}{c} \epsilon_{2{\scriptscriptstyle\perp}}\beta_2(u+i\alpha_2vw_{2e})/\epsilon \\ \beta_2(-iw_{2o}+\alpha_2uv) \\ \alpha_2\beta_2(v^2+\epsilon_{2{\scriptscriptstyle\perp}}\kappa^2) \end{array}\right), \\
\mathbf{m}_{B2}=\frac{1}{\kappa\sqrt{\epsilon\mu}\,\sqrt{u^2+v^2}}\left(\begin{array}{c} \epsilon_{2{\scriptscriptstyle\perp}}\delta_2(-iu+\gamma_2vw_{2e})/\epsilon \\ -\delta_2(w_{2o}+i\gamma_2uv) \\ -i\gamma_2\delta_2(v^2+\epsilon_{2{\scriptscriptstyle\perp}}\kappa^2) \end{array}\right). \label{mB2}
\end{gather}
The reflection operator $\mathbf{R}_2$ at Plate~2 for a general plane wave projects the wave to its components in the directions (\ref{nE}) and (\ref{nB2}), reflects these components according to (\ref{AE})--(\ref{mB2}), and adds the reflected components; it is therefore given by
\begin{equation} \label{R2}
\mathbf{R}_2=-(\mathbf{n}_E+\mathbf{m}_{E2})\otimes\mathbf{n}_E-(\mathbf{n}_{B2}+\mathbf{m}_{B2})\otimes\mathbf{n}_{B2}.
\end{equation}

\subsection{Reflection operator for Plate~1}
The reflection operator $\mathbf{R}_1$ for Plate~1 can of course be obtained from $\mathbf{R}_2$. At Plate~1 the incident wave has wave vector $\mathbf{k_-}$ rather than $\mathbf{k_+}$, but this just means $w\rightarrow -w$ compared to Plate~2. The $E$-polarization direction of the incident wave is therefore still given by (\ref{nE}), but the $B$-polarization direction $\mathbf{n}_{B1}$ is (\ref{nB2}) with $w\rightarrow -w$
\begin{equation} 
\mathbf{n}_{B1}=\frac{1}{\kappa\sqrt{\epsilon\mu}\,\sqrt{u^2+v^2}}\left(\begin{array}{c} i(u^2+v^2) \\ -uw \\ -vw \end{array}\right). \label{nB1}
\end{equation}
The only non-trivial issue is that the optical axis does not now lie in the $z$-direction. This is handled by changing the coordinate axes so that the optical axis lies along the new $z$-direction. The required passive coordinate transformation is a rotation about the $x$-axis by an angle $\theta$; this transforms vector components with the matrix $\bm{\Lambda}^{-1}=\bm{\Lambda}^T$, where $\bm{\Lambda}$ is given by (\ref{lambda}). We denote quantities in the rotated basis by primes, so that the components of the incident wave vector $\mathbf{k_-}$ in this basis are
\begin{gather}
{\mathbf{{k}'}_{\!\!-}}=(-iw,u',v'), \\
u'=u\cos\theta -v\sin\theta, \quad
v'=v\cos\theta +u\sin\theta.
\end{gather}
It is clear from (\ref{k2o})--(\ref{k2e}) that in the new frame the (left-moving) ordinary and extraordinary waves in Plate~1 have wave vectors 
\begin{gather}
{\mathbf{k}'}_{\!\!1o}=(-iw_{1o},u',v'), \quad w_{1o}=\sqrt{{u'}^2+{v'}^2+\epsilon_{1{\scriptscriptstyle\perp}}\kappa^2}, \label{k1o} \\
{\mathbf{k}'}_{\!\!1e}=(-iw_{1e},u',v'), \  w_{1e}=\sqrt{{u'}^2+{v'}^2\epsilon_{1{\scriptscriptstyle\parallel}}/\epsilon_{1{\scriptscriptstyle\perp}}+\epsilon_{1{\scriptscriptstyle\parallel}}\kappa^2}. \label{k1e}
\end{gather}
The reflection operator in the rotated frame, ${\mathbf{R}}'_1$, is now obtained from  $\mathbf{R}_2$ by obvious replacements. We then find the reflection operator in the frame of Fig.~\ref{fig1} from
\begin{equation}
\mathbf{R}_1=\bm{\Lambda}{\mathbf{R}}'_1\bm{\Lambda}^{T}.
\end{equation}
The result is
\begin{equation} \label{R1}
\mathbf{R}_1=-(\mathbf{n}_E+\mathbf{m}_{E1})\otimes\mathbf{n}_E-(\mathbf{n}_{B1}+\mathbf{m}_{B1})\otimes\mathbf{n}_{B1},
\end{equation}
where
\begin{gather}
\mathbf{m}_{E1}=\frac{1}{\sqrt{u^2+v^2}}\,\bm{\Lambda}\!\left(\begin{array}{c} \epsilon_{1{\scriptscriptstyle\perp}}\beta_1(u'-i\alpha_1v'w_{1e})/\epsilon \\ \beta_1(iw_{1o}+\alpha_2u'v') \\ \alpha_1\beta_1({v'}^2+\epsilon_{1{\scriptscriptstyle\perp}}\kappa^2) \end{array}\right), \\
\mathbf{m}_{B1}=\frac{1}{\kappa\sqrt{\epsilon\mu}\,\sqrt{u^2+v^2}}\,\bm{\Lambda}\!\left(\begin{array}{c} \epsilon_{1{\scriptscriptstyle\perp}}\delta_1(-iu'-\gamma_1v'w_{1e})/\epsilon \\ -\delta_1(-w_{1o}+i\gamma_1u'v') \\ -i\gamma_1\delta_1({v'}^2+\epsilon_{1{\scriptscriptstyle\perp}}\kappa^2) \end{array}\right),
\end{gather}
and $\alpha_1$, $\beta_1$, $\gamma_1$, $\delta_1$ are the Plate~1 versions of (\ref{al2})--(\ref{be2}) in the rotated frame:
\begin{widetext}
\begin{gather} 
\alpha_1=-\frac{iu'(\epsilon w_{1o}+\epsilon_{1{\scriptscriptstyle\perp}}w)}{v'(\epsilon w_{1o}^2+\epsilon_{1{\scriptscriptstyle\perp}}w_{1e}w)}, \qquad
\gamma_1=\frac{iv'w_{1o}(w+\mu w_{1o})}{\epsilon_{1{\scriptscriptstyle\perp}}u'\kappa^2(w+\mu w_{1e})}.   \\
\beta_1=\frac{-2i\epsilon v'w}{\epsilon w_{1o}w+\epsilon_{1{\scriptscriptstyle\perp}}{u'}^2-iu'v'\alpha_1(\epsilon w+\epsilon_{1{\scriptscriptstyle\perp}}w_{1e})+\epsilon\mu({v'}^2+\epsilon_{1{\scriptscriptstyle\perp}}\kappa^2)}, \quad
\delta_1=\frac{2\epsilon w({u'}^2+{v'}^2)}{\epsilon u'w_{1o}-i\epsilon\gamma_1v'w_{1o}^2+\epsilon_{1{\scriptscriptstyle\perp}}w(u'-i\gamma_1v'w_{1e})}. 
\end{gather}
\end{widetext}

With the reflection operators determined, we now have the solution for the Green tensor from (\ref{soln}). Equation~(\ref{soln}) contains inverse matrices and we evaluate these before using the Green tensor to compute the stress. This results in a lengthy expression for the Green tensor which we will not reproduce here. It is clear however from the form of the reflection operators (\ref{R2}) and (\ref{R1}), together with the geometrical-series expansions of the inverse matrices in (\ref{soln}), that the Green tensor is largely constructed from scalar products of the vectors $\mathbf{n}_E$, $\mathbf{n}_{B1}$, $\mathbf{n}_{B2}$, $\mathbf{m}_{E1}$, $\mathbf{m}_{E2}$, $\mathbf{m}_{B1}$ and $\mathbf{m}_{B2}$. Specifically, the following scalar products can be usefully employed to expand the Green tensor:
\begin{align}
\lambda_{1EE}=&\,(\mathbf{n}_E+\mathbf{m}_{E1})\cdot\mathbf{n}_E, \\
\lambda_{2EE}=&\,(\mathbf{n}_E+\mathbf{m}_{E2})\cdot\mathbf{n}_E, \\
\lambda_{1BB}=&\,(\mathbf{n}_{B1}+\mathbf{m}_{B1})\cdot\mathbf{n}_{B2}, \\
\lambda_{2BB}=&\,(\mathbf{n}_{B2}+\mathbf{m}_{B2})\cdot\mathbf{n}_{B1}, \\
\lambda_{1BE}=&\,\mathbf{m}_{B1}\cdot\mathbf{n}_E, \\
\lambda_{2BE}=&\,\mathbf{m}_{B2}\cdot\mathbf{n}_E,
\end{align} 
and these will appear in our solution for the Casimir stress and energy.

\section{\label{sec:stress}Stress tensor and energy}
Using the Green tensor we calculate the Casimir stress tensor $\bm{\sigma}$ from (\ref{stress}) and (\ref{EE})--(\ref{BB}).  It is first necessary to drop the bare Green tensor (\ref{bare}) in (\ref{soln}), as this gives the diverging zero-point stress in the absence of the plates~\cite{LL}. The component $\sigma_{xx}$ is the perpendicular Casimir force per unit area $F$ on the plates; we find
\begin{equation} \label{F}
F=\frac{\hbar c}{4\pi^3}\int_0^\infty \!\! d\kappa\int_{-\infty}^\infty \!\! du\int_{-\infty}^\infty \!\! dv\,w\!\left[\frac{Ae^{2aw}-2B}{e^{4aw}-Ae^{2aw}+B}\right],
\end{equation}
where
\begin{align}
A=&\,\lambda_{1EE}\lambda_{2EE}+\lambda_{1BB}\lambda_{2BB}-2\lambda_{1BE}\lambda_{2BE}, \\
B=&\,\left(\lambda_{1EE}\lambda_{1BB}+\lambda_{1BE}^2\right)\left(\lambda_{2EE}\lambda_{2BB}+\lambda_{2BE}^2\right).
\end{align}
If there is vacuum between the plates ($\epsilon=\mu=1$) the perpendicular force is always attractive. With a separating medium, however, even if $\mu=1$, the force can be attractive or repulsive~\cite{mun05}, as in the case of isotropic plates~\cite{dzy61}. Note that the dependence of $F$ on the plate separation $a$ is entirely visible in (\ref{F}), but the dependence on the orientation angle $\theta$ is hidden in the nested definitions that specify $A$ and $B$.

To compute the Casimir torque on the plates we require the Casimir energy $E$ ; this is related to the pressure $F$ and torque $Q$ by
\begin{align}
F=-\frac{\partial E}{\partial a}, \\
Q=-\frac{\partial E}{\partial \theta}. \label{Q}
\end{align}
We can therefore obtain $E$ by integrating (\ref{F}) with respect to $a$. This integration does not produce a function of integration depending on $\theta$ but not on $a$, nor a constant of integration, since $E$ must go to zero as $a\rightarrow\infty$. Our final result is therefore
\begin{equation} \label{E}
\begin{split}
E=&\frac{\hbar c}{4\pi^3}\int_0^\infty \!\! d\kappa\int_{-\infty}^\infty \!\! du\int_{-\infty}^\infty \!\! dv \\
&\times\left[2aw-\frac{1}{2}\ln \left(e^{4aw}-Ae^{2aw}+B\right)\right],
\end{split}
\end{equation}
which determines the torque through (\ref{Q}).

The previous exact analysis of this problem by Barash~\cite{bar78} found a very different looking formula for the Casimir energy. When $\mu=1$, the result (\ref{E}) must agree with the zero-temperature limit of Barash's expression. This agreement cannot be seen analytically but we have verified it numerically.

\section{Conclusions} 
We have calculated the Casimir energy for two parallel birefringent plates with optical axes parallel to the boundary, allowing for the presence of an isotropic medium between the plates. This is only the second exact treatment of the problem, and our result is in agreement with the previous solution~\cite{bar78}. Given the very different methods employed in the two calculations, this confirms the prediction of Casimir torque, an intriguing phenomenon that may be within reach of experiment~\cite{mun05}.  

\acknowledgments
We thank Y.\ S.\ Barash for a crucial communication. This research is supported by the Leverhulme Trust, the Max Planck Society and a Wolfson Research Merit Award of the Royal Society.

\end{document}